\documentclass[prl,twocolumn,showpacs,amsmath,amssymb]{revtex4}

\usepackage{amsmath,amssymb} 
\usepackage{bm}
\usepackage[dvips]{graphicx}
\usepackage{epsfig}     
\usepackage{dcolumn}
\usepackage{bm}
\newcommand{\ket}[1]{\vert{#1}\rangle}

\begin{document}

\title{Implementation of Grover's Quantum Search Algorithm \\
in a Scalable System}
\author{K.-A. Brickman, P. C . Haljan, P. J. Lee, M. Acton, L. Deslauriers, and C. Monroe}
\affiliation{FOCUS Center and Department of Physics, University of Michigan, Ann Arbor, MI  48109}
\date{\today}
\begin{abstract}
We report the implementation of Grover's quantum search algorithm in the scalable 
system of trapped atomic ion quantum bits.  Any one of four possible states of a two-qubit memory 
is marked, and following a single query of the search space, the marked element is 
successfully recovered with an average probability of 60(2)\%.  This exceeds the performance 
of any possible classical search algorithm, which can only succeed with a 
maximum average probability of 50\%.  
\end{abstract}
\maketitle

Quantum computers promise dramatic speedup over conventional computers in some applications owing to the power of entangled superpositions \cite{nielsen}.  Among the best-known quantum applications is Grover's search algorithm, which can search an unsorted database quadratically faster than any known classical search \cite{grover97a}. A common analogy for this searching algorithm is the problem of finding a person's name in a phone book given only their phone number \cite{brassard97}:  for $N$ entries in the phonebook, this requires of order $N$ queries.  However, if the correlation between name and phone number is encoded with quantum bits, the name can be found after only about $\sqrt{N}$ queries.  While Grover's algorithm does not attain the exponential speedup of Shor's quantum factoring algorithm \cite{shor97}, it may be more versatile,  by providing quadratic gains for almost any quantum algorithm \cite{grover98} or accelerating NP-complete problems through exhaustive searches over  possible solutions \cite{cerf00}.  

We implement the Grover search algorithm over a space of N=4 elements using two trapped atomic ion qubits \cite{feng01, hill04}.
Grover's algorithm has been implemented with ensembles of molecules using nuclear magnetic resonance \cite{chuang98, jones98, anwar04}, 
with states of light using linear optical techniques \cite{kwiat00, walther05}, and with Rydberg states
within individual atoms \cite{ahn00}.  None of these systems are scalable however, as they require exponential resources as the number of qubits grows.  
The implementation of Grover's algorithm reported here complements the repertoire of multi-qubit quantum algorithms recently demonstrated in the scalable system of trapped atomic ions \cite{riebe05, barrett05, schaetz04, chiaverini05}.  Unlike these earlier ion trap demonstrations, we use magnetically-insensitive ``clock state" qubits and particular entangling gates that are uniquely suited to such qubits while remaining insensitive to external phase drifts between gates \cite{haljan05a, haljan05, lee05}.

\begin{figure}[htbp]
\begin{center}
\includegraphics[width=1.0\columnwidth,keepaspectratio]{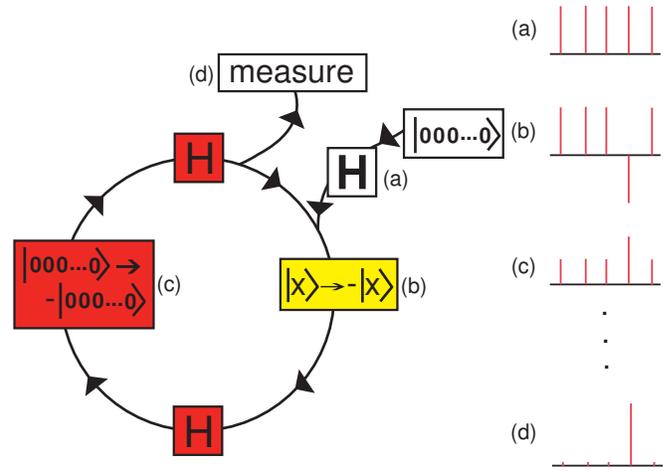}
\caption {Schematic diagram of Grover's quantum search algorithm over a space of
$n$ qubits ($N=2^n$ entries).  Initially, all qubits are prepared in the state $\ket{000...0}$.  (a) A global Hadamard gate prepares an equal superposition of all states.
(b) The oracle (shaded in yellow) flips the sign of the amplitude corresponding to
the marked element, represented by the $n-$bit binary number $\bf{x}$.  
(c) Two global Hadamard gates surround an additional phase gate (shaded in red)
that flips the sign of the amplitude corresponding to the initial state $\ket{000...0}$, 
amplifying the weight of the marked state.  
Steps (b) and (c) are repeated in sequence a prescribed number of times, and finally (d) the qubits are measured.
An example of the distributions of quantum amplitudes at each stage are depicted at the right. }
\label{fig:grover}
\end{center}
\end{figure}

At the heart of Grover's algorithm is the ``oracle query," which quickly checks if a proposed input $``x"$ is a solution to the search problem.  The oracle marks a particular component of a quantum superposition by flipping the sign of its amplitude.  Following the oracle, a number of quantum operations 
amplify the weighting of the marked state independent of which state is marked (see Fig. \ref{fig:grover}).  After many iterations of this query/amplification process, the marked state accumulates
nearly all of the weight and is revealed following a measurement.  
The required number of queries is the integer closest to $\frac{\pi}{4sin^{-1}(N^{-1/2})}$-$\frac{1}{2}$  \cite{boyer98}.  
For $N \gg 1$, the marked element would thus appear with high probability after approximately $\pi \sqrt{N}/4$ iterations, and for the special case of $N=4$ elements, a single query would provide the marked element with unit probability.  Classically, a single query of a 4-element search space followed by a guess can only result in a successful outcome with 50\% probability.

\begin{figure}[htbp]
\begin{center}
\includegraphics[width=1.0\columnwidth,keepaspectratio]{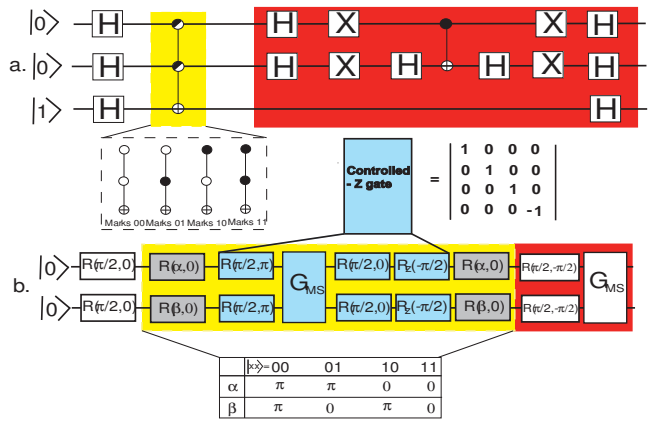} 
\caption{Quantum circuit to implement Grover's searching algorithm for N=4 entries \cite{nielsen}.  (a.) Theoretical circuit using a third ancilla bit and standard gates including the Haddamard gate (H), the generalized Toffoli gate, a bit flip,X, and a controlled-NOT gate.  The Toffoli gate implements the oracle (shaded in yellow), where the scheme to mark each of the four possibilities is shown below the circuit.  The remainder of the circuit (shaded in red) amplifies the weighting of the marked state.  (b.) The experimental circuit to implement the algorithm for n=2 qubits, where R($\theta,\phi$) is a rotation on the Bloch sphere \cite{grovercircuit}, $R_{z}$($\phi$) is a phase rotation about the $\hat{z}$-axis, and $G_{MS}$ is the M$\textnormal{\o}$lmer-S$\textnormal{\o}$rensen  entangling gate.  The yellow shaded box identifies the oracle, where the value of the variables $\alpha$ and $\beta$ (given in the table), determine which state is marked \cite{grovercircuit2}.  The remainder of the circuit (shaded in red) amplifies the weighting of the marked state.}
\label{fig:circuit}
\end{center}
\end{figure}

A standard quantum circuit for the Grover search algorithm for N=4 entries is shown in Figure \ref{fig:circuit}(a) \cite{nielsen}.  This scheme uses a third ancilla bit which marks one of the database elements through a Toffoli gate that effectively flips the sign of the marked element if and only if the two bit input is a solution to the problem.  The oracle scheme to mark each of the four possibilities is shown below the circuit.  The remainder of the circuit (shaded in dark red) amplifies the weighting of the marked state, with the operations between the Hadamard gates flipping the sign of the amplitude of the $\ket{00}$ state.

Fig. \ref{fig:circuit}(b) shows the experimental implementation of the algorithm for N=4 search elements.  The M$\textnormal{\o}$lmer-S$\textnormal{\o}$rensen (M-S) entangling gate \cite{sorensen99} is adapted to the algorithm for both the marking and state amplification steps and is accompanied by a variety of single qubit rotations.  This circuit is identical in function to the circuit of Fig. \ref{fig:circuit}(a) but does not use an ancilla qubit \cite{dodd03}.  First, each qubit is initialized to the $\ket{0}$ state through optical pumping techniques.  Next, an equal superposition of all qubit states is prepared with a global single qubit rotation of $\pi$/2, replacing the Hadamard gates of Figs \ref{fig:grover} and \ref{fig:circuit}(a).  The oracle function (shaded in yellow) then marks one of the four possible states $\ket{xx}$ by flipping its sign.  This is accomplished with a controlled-z phase gate (shaded in blue) containing the M-S gate, that flips the sign of the $\ket{11}$ state.  The phase gate is surrounded by rotations (shaded in gray) that swap the marked state $\ket{xx}$ with the $\ket{11}$ state, depending on the settings of the of the rotation angles depicted in the table at the bottom of Fig \ref{fig:circuit}(b).  After the oracle query, a global rotation of $\pi$/2 (with a phase of -$\pi$/2 relative to the initial rotation) followed by a second application of the M-S gate amplifies the weighting of the marked state $\ket{xx}$ (operations shaded in red).  

The experiment is performed with two $^{111}Cd^{+}$ ions confined  in a three-layer linear ion trap with axial frequency $\omega_{z}$/$2\pi$=2.0MHz \cite{Deslauriers2004a, haljan05, haljan05a}.  The $S_{1/2}$ ground state hyperfine levels $\ket{F=0, m_{f}=0}$ (denoted by $\ket{0}$) and $\ket{F=1, m_{f}=0}$ (denoted by $\ket{1}$), separated in frequency by $\omega_{0}/2\pi$= 14.5 GHz, serve as qubit levels.  These ``clock" states are relatively insensitive to magnetic field noise, obviating the need for ``spin echo" pulses in the implementation of the algorithm \cite{barrett05}.  A variety of optical and microwave sources are used to control, entangle, and measure the qubits.  In particular, pairs of off-resonant laser beams with a beatnote near $\omega_{0}$ drive stimulated Raman transitions that couple the qubits to their collective motion and give rise to entangling quantum gates \cite{lee2003a, haljan05}.  Both the center-of-mass and stretch modes of axial motion are initialized to near the ground state via stimulated Raman cooling, with average vibrational numbers of $\overline{n}$=0.28 and 0.12, respectively \cite{Deslauriers2004a}.  Single qubit rotations are accomplished through resonant microwave pulses (tuned to near $\omega_{0}$) and composite pulse sequences involving a single tightly-focused off-resonant laser.  The microwaves are applied to both qubits simultaneously and provide global single qubit rotations.  The off-resonant laser, which addresses one ion more strongly than the other, creates a differential AC Stark shift between the two qubits, imparting a relative phase shift of $\pi$ between the two qubits.  Combined with appropriate microwave pulses, this allows arbitrary single qubit rotations.  Finally, each qubit is detected with greater than 97$\%$ efficiency by uniformly illuminating the ions with resonant laser radiation and observing the ion fluorescence on an intensified CCD camera.

The M$\textnormal{\o}$lmer-S$\textnormal{\o}$rensen gate directly entangles the clock state qubits and is insensitive to the relative optical phase of the Raman laser beams between gates\cite{haljan05a, haljan05, lee2003a}.  This is an important consideration when multiple entangling gates are implemented because it suppresses decoherence from magnetic fields and optical phase noise that may fluctuate from gate to gate.  The M-S entangling gate is realized by applying multiple sets of Raman beatnotes to the ions, simultaneously driving the first lower and upper motional sidebands for a particular duration.  This entangles the spin states via their collective motional mode, in this case the stretch mode \cite{sorensen99}.  The evolution of the four basis states is given by:

\begin{description}
\item $\ket{0} \ket{0} \rightarrow \ket{0} \ket{0}-i\ket{1} \ket{1}$
\item $\ket{1} \ket{1} \rightarrow \ket{1} \ket{1}-i\ket{0} \ket{0}$
\item $\ket{0} \ket{1} \rightarrow \ket{0} \ket{1}-i\ket{1} \ket{0}$
\item $\ket{1} \ket{0} \rightarrow \ket{1} \ket{0}-i\ket{0} \ket{1}$.
\end{description}

\noindent The desired states are produced with a fidelity of approximately 80\%.  There are additional phases not included in the above equation pertaining to the ion-ion spacing, the phase of the oscillator that defines the Raman beam beatnote, and Stark shifts from the applied Raman beams \cite{haljan05a, lee05}.  These phases are set to zero for the present case since they are fixed before the experiment is run by first synchronizing the phase of the entangling gate with the phase of microwave $\pi/2$ pulses\cite{haljan05a}.  Then the phases of the two entangling gates used in the experiment are  synchronized to each other through a Ramsey experiment.  Note that phase noise from the Raman beam path fluctuations is supressed due to the spectral arrangement of the Raman sidebands \cite{lee2003a, haljan05}.   The timescale for each operation in the algorithm is as follows: $10\mu$s for a global microwave rotation, $20\mu$s for a differential single qubit rotation, and $140\mu$s for the M$\textnormal{\o}$lmer-S$\textnormal{\o}$rensen two qubit entangling gate, giving a total of $\sim$ $380\mu$s to complete the 20 pulses that form the algorithm.

\begin{figure}[htbp]
\begin{center}
\includegraphics[width=1.0\columnwidth,keepaspectratio]{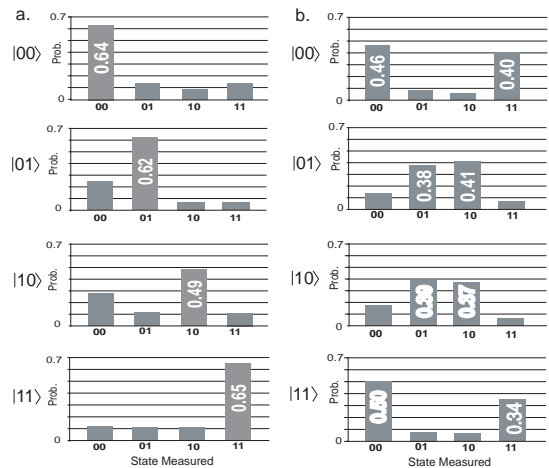} 
\caption{(a.) Output of the algorithm.  The conditional probability of measuring each of four output states given one was marked is shown in sequence from top to bottom $\ket{00}, \ket{01}, \ket{10}, \ket{11}$. Each of the four data sets shows the distribution of measurements averaged over 500 trials.  The marked state was recovered on average with 60(2)\%, compared to unit probability for the ideal quantum algorithm and 50\% for the best possible classical algorithm.  (b.)  Output of the algorithm without the final entangling gate.  This shows that the fidelity of the oracle is about $80\%$.  Each of the four data sets was also averaged over 500 trials.  The experimental average to recover the marked state is 41(2)\% with the theoretical limit of 50\%, both of which are less than the 60\% from (a).  The quoted errors are statistical.
}
\label{fig:data}
\end{center}
\end{figure}

There are several approaches to gauging the performance of the algorithm implementation.  One method is to compare the algorithm's success at recovering the marked state with the best that can be achieved classically.  The classical counterpart is a simple shell game: suppose a marble is hidden under one of four shells, and after a single query the location of the marble is guessed.  Under these conditions, the best classical approach gives an average probability of success $P_{cl}=1/4+3/4(1/3)=0.50$, because 1/4 of the time the query will give the correct location of the marble while 3/4 of the time a guess must be made amongst the three remaining choices each with 1/3 probability of choosing the correct location.  If Grover's algorithm is used, the answer to the single query would result in a 100$\%$ success rate at `guessing' the marble's location.  As can be seen in figure \ref{fig:data}(a) the marked state is recovered with an averaged probability over the four markings of 60(2)\%, surpassing the classical limit of 50\%.

It is interesting to consider the output of the algorithm when the final entangling gate used for state amplification is omitted.  This situation shows how well the algorithm can do with only single qubit rotations outside the oracle.  This scenario lies between the classical and quantum searches described above since entanglement is not used outside the oracle but quantum superpositions are used to find the marked element.  In this case it can be shown that quantum mechanics without entanglement can do no better than what can be achieved with classical means: both methods have the outcome of finding the marked state with only $50\%$ probability, assuming a perfect oracle.  In addition, this diagnostic allows the performance of the oracle itself to be characterized.  The rotations following the oracle convert the marked state into one of four Bell states each of which yields a  maximum probability of 50\% to recover the marked state.  Figure \ref{fig:data}b shows that the marked state is recovered with an average of 42(1)\% probability, implying the oracle itself has a fidelity of roughly $80\%$. 

The above figures of merit focus on the mean success probability and neglect the information content inherent in the distributions of figure \ref{fig:data}a.  The mutual information between the marking of the state and the measurement can be used to characterize this correlation and hence is another measure of the algorithm's success \cite{nielsen} (for other figures of merit see ref. \cite{dodd03}).  The mutual information measures how much information two random variables, x, the measurement, and y, the marking, have in common.  It is defined by: H(x:y)=$H(x)+ H(y)- H(x,y)$, where $H(x,y)$=-$\sum_{x,y}$ p(x,y) log$_{2}$ p(x,y) is the joint Shannon entropy between the two distributions, $p(x,y)=p(x)p(y|x)$ is the joint probability distribution of x and y, and $p(y|x)$ is the conditional probability of y having been marked given that x was measured.  $H(x)$=-$\sum_{x}$p(x) log$_{2}$ p(x) and $H(y)$=-$\sum_{y}$p(y) log$_{2}$ p(y) are the Shannon entropies of the individual variables.  Classically the mutual information acquired after a single query of the oracle is  H(x:y)=0.25log$_{2}$(0.25)-0.75log$_{2}$(0.75)=0.81 bits, meaning, on average, 0.81 bits of information are gained upon measurement.  The ideal quantum algorithm would yield two bits of information upon measurement.  For the data in fig.\ref{fig:data}a the mutual information is 0.44, so on average only about a half a bit of information is gained.  Even though  less information is gained per measurement than the classical case, the $\it{probability}$ of finding the marked state in the experiment still exceeds the classical limit.

As the data in fig. \ref{fig:data}a shows, the marked state is not recovered with unit probability.  A large part of this infidelity is due to the M{\o}lmer-S{\o}rensen entangling gate.  Each instance of the M{\o}lmer-S{\o}rensen gate has a fidelity of about 80\%, and since there are two such gates in the algorithm, overall fidelities of approximately 60\% are expected.  The main sources of decoherence during the gate are spontaneous emission from off-resonant coupling to the excited state and fluctuating AC Stark shifts from the Raman beams that drive the entangling gate\cite{haljan05}.  Both of these induced decoherence sources can be suppressed by increasing the detuning of the Raman beams from the excited state, at the expense of slowing the gate.  We choose the detuning to strike a balance between these induced decoherence sources and other slowly varying noise sources, such as motional heating \cite{Turchette2000, Deslauriers2004a}, fluctuating magnetic fields, and microwave oscillator phase drifts.  Additional power in the Raman laser beams accompanied by larger detunings could suppress decoherence from spontaneous emission and AC Stark shifts while maintaining a reasonable gate speed (see \cite{haljan05} for more details).  Fluctuating AC Stark shifts during the differential single qubit rotations due to technical intensity fluctuations and beam pointing instabilities add infidelities to the experiment on the order of 5-10$\%$.

For Grover's algorithm to be useful it needs to extend beyond a few qubits.  Using a quantum circuit similar to fig.\ref{fig:circuit}a, an $\it{n}$-qubit Grover algorithm can be implemented with $\it{n}$-qubit Toffoli gates, a series of two qubit gates, and single qubit rotations.  It has been shown that an $\it{n}$-qubit Toffoli gate can be constructed with single qubit gates and controlled-NOT gates with order $\it{n}$ basic operations \cite{barenco95}.  A controlled-NOT gate can be constructed from the M-S entangling gate through the following sequence \cite{sorensen99}: 
[$R_{2}(\pi/2,0),R_{1}(\pi/2,\pi),R_{2}(\pi/2,\pi),G_{MS},R_{1}(\pi/2,0),\\ R_{2}(\pi/2,0),R_{z1}(-\pi/2),R_{z2}(-\pi/2),R_{2}(\pi/2,-\pi),R_{z1}(\pi)$], where R$_{i=1,2}$($\theta,\phi$) is a rotation of ion $\it{i}$  by angle $\theta$ and phase $\phi$, $R_{zi}$($\phi$) is a z-rotation of ion $\it{i}$ by angle $\phi$, and G$_{MS}$ is the M$\textnormal{\o}$lmer-S$\textnormal{\o}$rensen entangling gate.  Since the ion system is scalable to a large number of qubits it is feasible to construct an efficient $\it{n}$-qubit Grover algorithm where each iteration scales polynomially with $\it{n}$.  In this case, the isolation of individual ions could be accomplished through tight focusing of laser beams\cite{schmidt-kaler03} or the shuttling of ions between separated trap zones \cite{rowe02, kielpinski02, barrett05, schaetz04, chiaverini05}.

This work is supported by the National Security Agency and Advanced Research and Development Activity under Army Research Office contract W911NF-04-1-0234, and the National Science Foundation Information Technology Research Program.

\end{document}
}

\bibitem[{zrot()}]{grovercircuit2}
\bibinfo{note}{In the experiment the first set of rotations (shaded in gray) are accompanied by extra z-rotations.  These additional phases do not affect the outcome of the oracle and are suppressed in the figure for simplicity.}

\end{thebibliography}


\begin{thebibliography}{30}
\bibitem[{zrot()}]{grovercircuit2}
\bibinfo{note}{In the experiment the first set of rotations (shaded in gray) are accompanied by extra z-rotations.  These additional phases do not affect the outcome of the oracle and are suppressed in the figure for simplicity.}
\end{thebibliography}
\end{document}